\documentclass[12pt]{iopart}
\usepackage{iopams} 
\usepackage[dvips]{graphicx}
\usepackage{cite}

\begin{document}
\title{Surface parameters of ferritic iron-rich Fe-Cr alloy}

\author{S Sch\"onecker$^1$, S K Kwon$^2$, B Johansson$^{1,3}$ and L Vitos$^{1,3,4}$}
\address{$^1$ Applied Materials Physics, Department of Materials Science and Engineering, Royal Institute of Technology, Stockholm SE-10044, Sweden}
\address{$^2$ Graduate Institute of Ferrous Technology, Pohang University of Science and Technology, Pohang 790-784, Korea}
\address{$^3$ Department of Physics and Astronomy, Division of Materials Theory, Uppsala University, Box 516, SE-75120 Uppsala, Sweden}
\address{$^4$ Research Institute for Solid State Physics and Optics, Wigner Research Center for Physics, Budapest H-1525, P.O. Box 49, Hungary}
\ead{stesch@kth.se}

\date{\today}

\begin{abstract}
Using first-principles density functional theory in the implementation of the exact muffin-tin orbitals method and the coherent potential approximation, we studied the surface energy and the surface stress of the thermodynamically most stable surface facet (100) of the homogeneous disordered body-centred cubic iron-chromium system in the concentration interval up to 20 at.\% Cr. 
For the low-index surface facets of Fe and Cr, the surface energy of Cr is slightly larger than the one of Fe, while the surface stress of Cr is considerably smaller than the one of Fe. We find that Cr addition to Fe generally increases the surface energy of the Fe-Cr alloy, however, an increase of the bulk amount of Cr also increases the surface stress. As a result of this unexpected trend, the (100) surface of Fe-Cr becomes more stable against reconstruction with increasing Cr concentration. We show that the observed trends are of magnetic origin. In addition to the homogeneous alloy case, we also investigated the impact of surface segregation on both surface parameters.
\end{abstract}
\maketitle
\submitto{\JPCM}
\section{\label{sec:intro}Introduction}

In recent years, iron rich Fe-Cr alloys, as the basis of ferritic (body-centred cubic (bcc)-based) stainless steel, have attracted much scientific attention for their potential use in the next generation fission and prospective fusion reactors~\cite{Cook:2006}. Employed as first wall and blanket material, or fuel cladding, this steel must withstand neutron-induced radiation damage~\cite{Klueh:2007}, e.g., swelling and void formation. Moderate Cr addition to bcc Fe in the order of 10\,\% most beneficially improves its swelling and irradiation creep behaviour~\cite{Garner:2000,Little:1979}.
Chromium substitution is further known to influence other mechanical properties such as the ductile-brittle transition temperature and radiation-induced hardening~\cite{Kohyama:1996}. 
The experimental corrosion resistance of ferritic steel in an oxidising environment improves drastically if the bulk alloy concentration exceeds 9-13 weight percent Cr and this self-healing protection attributed to the formation of a chromium oxide scale~\cite{Wranglen:1985,Khanna:2002}. There is, however, a limit to the amount of Cr that may be added to steels as the beneficial low corrosion rate is shadowed by the enhanced precipitation of intermetallic phases which often degrade the mechanical properties of ferritic steel. 
Moreover, it has also been recognised that alloying with additional elements further increases the pitting and crevice corrosion resistance in certain aggressive environments, e.g., molybdenum in chloride environments~\cite{Gunn:1997}. 

The surface of the bcc Fe-Cr system has often been used as the prototype reference system to study the behaviour of surfaces of ferritic stainless steels. There is a great deal of phenomenological modelling on the passivity of stainless steels~\cite{Williams:1991,Sieradzki:1986,Burstein:1984}, while first principles investigations of surfaces focused on the atomic level behaviour involving surface segregation and the surface magnetic structure for the technologically relevant Fe-rich Fe-Cr alloys~\cite{Kiejna:2008,Ropo:2007,Ropo:2011,Klaver:2006,Ponomareva:2007,Ruban:1999,Levesque:2012,Nonas:1998,Geng:2003}.
Understanding the aforementioned threshold behaviour of the corrosion resistance and the particular role of Cr in the segregation process has been a primary target for modelling. Experimental evidence indicates Cr enrichment at the surface at high temperatures and bulk Cr concentrations larger than 13\,at.\%~\cite{Dowben:1983,Lince:1992,Suzuki:1996,Leygraf:1974}. 
Recent ab initio calculations indeed predicted a sharp transition from Cr-free to Cr enriched surfaces at around 9\,at.\% Cr in the bulk alloy~\cite{Ropo:2007,Ropo:2011,Kiejna:2008}, attributed to complex magnetic interactions between ferromagnetically interacting Fe and anti-ferromagnetically interacting Cr species, which are likely to govern many essential characteristics of the Fe-Cr phase diagram below the Curie temperature~\cite{Ackland:2006}.

Not much is known about the ferritic Fe-Cr system concerning two essential macroscopic parameters that describe the thermodynamic properties of its crystalline surface: surface energy and surface stress. That is surprising since the significance of stress and strain effects on surface physics has been widely discussed~\cite{Mueller:2004,Ibach:1997,Ibach:1997E,Haiss:2001}.
The equilibrium shape of mesoscopic crystals is the one that minimises its surface free energy~\cite{Wulff:1901}. The surface energy is further of eminent relevance in connection to faceting, roughening, crystal growth phenomena at surfaces and has been discussed in relation to segregation. There has been an increasing experimental and theoretical activity to understand the importance of stress on many physical properties associated with surface relaxation and reconstruction~\cite{Bach:1997,Olivier:2003}, segregation~\cite{Wynblatt:1977}, surface adsorption~\cite{Ibach:2004}, and its role in bottom-up self-organisation and surface melting~\cite{Mueller:2004}. 
On the other hand, a theoretical study of surface parameters of crystalline surfaces of the individual elements, Fe and Cr, was subject to a number of publications~\cite{Alden:1992,Punkkinen:2011,Punkkinen:2011b,BlancoRey:2010,Ossowski:2008}.
According to the general expectation, surface stress of clean surfaces is tensile due to the increased electron density within the surface layer. However, it was demonstrated recently~\cite{Punkkinen:2011,Punkkinen:2011b} that magnetism can overwrite this picture leading to exceptional, compressive surface stresses as predicted for the thermodynamically stable surfaces of bcc Cr and cubic Mn, though not in the case of bcc Fe.
It is hence worthwhile to investigate the surface of the bcc Fe-Cr system to gain information on its essential surface parameters keeping in mind the presence of complex magnetic interactions in this alloy system that may alter our expectation on their behaviours.
In this context, it is important to mention the particular role of ab initio calculations in the determination of surface parameters since experimental methods to determine their absolute value often lack reliability and accuracy~\cite{Ibach:1997,Ibach:1997E,Lamber:1995,Wasserman:1970,Wasserman:1972,Solliard:1985,Mays:1968,Metois:2004}.

This work deals with an ab initio determination of surface parameters for the thermodynamically stable surface of bcc Fe-Cr ((100) facet) in the concentration range of 0-20\,at.\% Cr.
The paper is organised a follows: In Section~\ref{sec:surfparam} we briefly overview the theory of surface energy and surface stress. Numerical details of our computation are presented in Section~\ref{sec:esc}. We discuss our results in Section~\ref{sec:results} for two different surface chemistries: a perfectly truncated bulk system without spacial concentration dependence and a system involving surface segregation. The reason for the latter is to account for the observed transition from Cr-free to Cr enriched surfaces around the aforementioned threshold bulk Cr concentration.

\section{\label{sec:surfparam}Surface parameters}

Surface energy and surface stress are two fundamental quantities to characterise the macroscopic properties of surfaces. Qualitatively, the scalar surface free energy, $\gamma$, was introduced by Gibbs as the reversible work per unit area to create a surface~\cite{Ibach:1997,Ibach:1997E}. The tensorial surface stress, $\tau_{ij}$, $i,j=\{x,y\}$, is the reversible work per unit area to stretch a surface elastically in the surface plane which is here assumed to lie in the $x$-$y$-plane.
If $\gamma<0$ for a particular surface of a solid, then this surface is unstable and the crystal fragments spontaneously. Hence $\gamma$ is positive for stable bulk systems. The components of $\tau$ may be both positive (tensile surface stress) or negative (compressive surface stress). Tensile surface stress on a surface favours smaller in-plane lattice constants than the bulk value while a surface with compressive surface stress favours a larger one.

In ab initio total energy calculations, the surface energy is usually computed as~\cite{Mueller:2004,Ibach:1997,Ibach:1997E},	
\begin{equation}
 \gamma =  \frac{E_{\rm surf} - E_{\rm bulk}}{A},
\label{eq:surfaceenergy}
\end{equation}
where $E_{\rm surf}$ and $E_{\rm bulk}$ specify the energy of two semi-infinite bulk systems and the infinite bulk system, respectively, normalised to the unit area $A$. 
Surface energies are conveniently extracted from slab calculations and different procedures were proposed to yield convergent numbers with the slab size~\cite{Needs:1987,Boettger:1994,Fiorentini:1996}. In the present work, we model the Fe-Cr system by considering two distinct subsystems: one that includes the surface (surface subsystem) and one without (bulk subsystem). 
Due to the periodic boundary conditions parallel to the surface, the size of the slab has to be converged with respect to the thickness of the slab only. 
Here, we follow essentially~\cite{Needs:1987} and derive both $E_{\rm surf}$ and $E_{\rm bulk}$ from slabs with the same thickness characterised by the total number of layers, $n$, taken as a multiple of the bulk equilibrium lattice parameter oriented normal to the surface plane. In case of the surface subsystem, the slab consists of an atomic part with thickness $n_{\rm m}$ and vacuum which is needed to decouple the two surfaces of the slab from another (across the vacuum). This surface-surface distance is denoted by $n_{\rm v}$. Since the slab representing the surface subsystem contains two equal surfaces, a factor of one half is added to \eref{eq:surfaceenergy} to yield the surface energy of one surface, viz.
\begin{equation}
 \gamma =  \frac{E^n_{\rm surf} - \frac{n_{\rm m}}{n}E^n_{\rm bulk}}{2A},
\label{eq:surfaceenergy:slabs}
\end{equation}
where $n=n_{\rm m}+n_{\rm v}$, and both $E^n_{\rm surf}$ and $E^n_{\rm bulk}$ refer to the total energy of the $n$-layer slab. $E^n_{\rm bulk}$ is scaled by a factor of $n_{\rm m}/n$ to the correct number of atoms. Only geometry relaxation in the direction perpendicular to the surface is allowed and may be included in $E^n_{\rm surf}$ as appropriate.

The surface stress tensor can be defined as the strain derivative of the surface energy in the Lagrangian coordinate system (surface area $A$ is standard state of strain)~\cite{Cahn:1980,Mueller:2004},
\begin{equation}
 \tau_{ij} = \left.\frac{\partial \gamma}{\partial \epsilon_{ij}}\right|_{\epsilon_{ij}=0},
 \label{eq:defstresslagr}
\end{equation}
where $\epsilon_{ij}$ denotes the components of the strain tensor specifying an elastic in-plane deformation of the surface. The $\tau_{ij}$'s are evaluated at the unstrained state ($\epsilon_{ij}=0$).
For a high-symmetry surface facet such as the bcc (100) facet, $\tau_{xx}=\tau_{yy}$, $\tau_{xy}=\tau_{yx}=0$, further assuming an isotropic distortion, $\epsilon_{xx}=\epsilon_{yy}=\epsilon$ (zero otherwise), and using the surface energy from \eref{eq:surfaceenergy:slabs}, we arrive at
 \begin{equation}
\tau\equiv\tau_{ii} = \left.\frac{1}{4A}\frac{\partial (E^{n}_{\rm surf}(\epsilon) - \frac{n_{\rm m}}{n} E^{n}_{\rm bulk}(\epsilon))}{\partial \epsilon}\right|_{\epsilon=0}.
\label{eq:surfstress:bcc100}
\end{equation}
A factor of $1/2$ appears due to the applied isotropic strain. The previous equation is conveniently~\cite{Kollar:2003,Punkkinen:2011,Punkkinen:2011b,Kadas:2006,Kwon:2005} used to compute the surface stress from the elastic energies of a surface subsystem and a bulk reference system employing slabs.

Numerically, the elastic energies in \eref{eq:surfstress:bcc100} are fitted to second order polynomial functions with fit coefficients $c$ and $d$ as a function of strain,
\numparts
\begin{eqnarray}
E^{n}_{\rm surf}(\epsilon)-E^n_{\rm surf} &=& c_{\rm surf} \epsilon + d_{\rm surf} \epsilon^2 \label{eq:fita} \\ 
\frac{n_{\rm m}}{n}\left(E^{n}_{\rm bulk}(\epsilon)-E^n_{\rm bulk}\right) & = & c_{\rm bulk} \epsilon + d_{\rm bulk} \epsilon^2.
\label{eq:fitb}
\end{eqnarray}

The reference energies, $E^n_{\rm surf/bulk}$, are the energies of the unstrained surface and bulk states. %, which may serve to compute $\gamma$ following \eref{eq:surfaceenergy:slabs}. 
Assuming the previous fit functions, we may express $\tau$ by
\endnumparts
\begin{equation}
 \tau = \frac{c_{\rm surf}-c_{\rm bulk}}{4A}.
\label{eq:surfstress:fit}
\end{equation}
Although in theory $c_{\rm bulk}\equiv 0$ for bulk in equilibrium, it may be finite in calculations due to numerical errors~\cite{Punkkinen:2011b}.

We conclude this section by mentioning the excess surface stress~\cite{Ibach:1997,Ibach:1997E,Cammarata:1992},  $|\tau-\gamma|$. A larger value of the excess surface stress, $|\tau-\gamma|$, indicates a higher tendency of a surface towards reconstruction~\cite{Cammarata:1992,Zolyomi:2009}.

\section{\label{sec:esc}Electronic structure calculations}

Total energy calculations within the framework of density functional theory (DFT) were done by means of the exact muffin tin orbitals (EMTO) method~\cite{Vitos:2007,Vitos:2001a,Vitos:2000}, which is a screened Korringa-Kohn-Rostoker type of method~\cite{Zabloudil:2005,Korringa:1947,Kohn:1954} and solves the Kohn-Sham equations in a Green's function formalism. This enables us to compute the electronic structure of substitutionally disordered alloys using the coherent-potential approximation (CPA)~\cite{Soven:1967,Gyorffy:1972,Faulkner:1982}. The one-electron potential is represented by optimised overlapping muffin-tin potential spheres describing more accurately the exact crystal-potential compared to conventional muffin-tins or non-overlapping spherical symmetry potentials~\cite{Vitos:2007,Zwierzycki:2009}. Combined with the full-charge density technique~\cite{Kollar:2000,Vitos:1994,Vitos:1997a,Vitos:1997b} for total energy calculations,~\cite{Vitos:2007,Vitos:2001a} the EMTO method has proven to yield reliable total energies and electronic structure in practise including the Fe-Cr system under consideration~\cite{Vitos:2001b,Ropo:2011,Ropo:2007,Ruban:2008,Razumovskiy:2011,Olsson:2003,Olsson:2006,Zhang:2009,Korzhavyi:2009}.
The CPA, being a single-cite approximation to the impurity problem, is a standard technique for electronic structure calculations in totally random alloys, suited for  the case of alloy components having similar sizes. Due to its single-site nature, screening corrections to the potential and the total energy must be taken into account which was done within the screened impurity model of Korzhavyi \etal~\cite{Korzhavyi:1995,Ruban:2002a,Ruban:2002b}. 
The CPA cannot directly treat atomic short-range order (ASRO), which was observed for the nearest neighbour coordination shell for the Fe-Cr system by means of neutron-diffusive scattering~\cite{Mirebeau:1984,Mirebeau:2010}. Accordingly, there is a tendency to form an ordered compound for Cr concentrations smaller than 11\,at.\% while larger Cr contents incline short-range clustering. Recent studies by M\"ossbauer spectroscopy reported that the ASRO inversion occurs at 6\,at.\% Cr~\cite{Dubiel:2011,Idczak:2012}. 
Inclusion of ASRO is, however, beyond the scope of this work. 
We note in this context that the surface energy and the surface stress are defined as excess quantities, i.e., contributions from the bulk part of the surface subsystem and the bulk subsystem are expected to cancel each other out (e.g., ASRO) and only surface effects survive. To the best of our knowledge, there is neither experimental nor theoretical evidence on ASRO at the surface of Fe-Cr.

Total energies in EMTO were obtained using the Perdew-Burke-Ernzerhof (PBE) parametrisation~\cite{Perdew:1996,Perdew:1996E} of the exchange-correlation energy functional, while self-consistent charge densities were computed in the local-density approximation in the parametrisation of~\cite{Perdew:1992}. Switching off gradient corrections to the exchange-correlation potential is justified since functionals in the generalised-gradient approximation are known to overestimate the magnetic moment of iron. This perturbative approach gives accurate total energies which was also tested in the case of Fe-Cr~\cite{Korzhavyi:2009,Asato:1999}. 
The EMTO partial waves were expanded into $s$, $p$, $d$, and $f$ orbitals, and the core states were recalculated at each iteration step. Integration over the Brillouin zone was done on a $k$-point grid of $15\times15\times1$ points in case of the surface subsystem. A single $k$-point along the short reciprocal lattice vector (corresponding to the direct lattice vector parallel to $z$) is sufficient since bands are dispersionless in this direction (ensured by the converged thickness
of vacuum). For the bulk reference system, the number of $k$-points in this direction was converged to the value of two. Increasing the $k$-point grid of the surface subsystem (bulk subsystem) to $20\times20\times1$ ($20\times20\times3$) showed that the total energies are converged at a level of $1\,\mathrm{meV}$/atom. The Green's function was evaluated for 16 complex points distributed exponentially on a semicircle including states below the Fermi level. 

For the particular case of Fe we also carried out DFT calculations with the full-potential local-orbital scheme, FPLO-9.01-35~\cite{Koepernik:1999} and PBE. The convergence of numerical parameters and the basis was carefully checked. Linear-tetrahedron integrations with Bl\"ochl corrections on a $12 \times 12 \times 1$ ($12 \times 12 \times 3$) mesh in the full Brillouin zone for the surface subsystem (bulk subsystem) were sufficient to converge the total energy at a level of
$1\,\mathrm{meV}$/atom compared to a $18 \times 18 \times 1$ ($18 \times 18 \times 5$) mesh. The valence basis of Fe comprised $3spd$, $4spd$, and $5s$ states.

Ferritic Fe-Cr crystallises in the bcc crystal structure, and the (100) surface was reported to be the most stable facet~\cite{Ropo:2007}. In case of EMTO, we modelled this surface by a slab with a converged thickness of 13 atomic layers (approximately $18\,{\rm \AA}$) and separated by vacuum with a thickness of $n_{\rm v}=7$ (approximately $10\,{\rm \AA}$). A somewhat larger surface subsystem with $n_{\rm m}=19$ and $n_{\rm v}=7$ was required to yield converged surface parameters in case of FPLO. The symmetry of the slabs includes a mirror plane parallel to the (identical) surfaces.

Surface geometries may differ from ideally truncated bulk crystals since relaxation and reconstruction may occur. Surface reconstruction on the close-packed surfaces of the transition metals is rather uncommon~\cite{Watson:1994}, however relaxation of the surface layer and of subsurface layers is frequently observed. 
Results of~\cite{Blonski:2007} and references therein indicate that layer relaxations of the most stable (110) and the second most stable (100) surface facets of Fe are minor and change the corresponding surface energies in the order of 1\,\% (cf.$\!$ literature values in table~\ref{table:results:Fe}).
Punkkinen \etal~recently compared the surface energy and the surface stress of the most stable surfaces facets of Fe and Cr for non-relaxed geometry with values of fully relaxed surface geometries~\cite{Punkkinen:2011}. 
Due to the enhanced surface magnetism in both systems, relaxations have only a minor effect and they were found not to markedly alter the surface energy and the surface stress. 
We expect for the same reasons that the surface geometry of the bcc (100) surface of Fe-Cr remains close to the truncated bulk one.
To further support this point, we computed the surface-layer relaxation of the bcc (100) surface of Fe and of Fe-Cr (up to 20\, at.$\%$ Cr) with EMTO. 
We found a relative change in length of the interlayer distance of $-2.2\,\%$ for Fe (inward relaxation) and an accompanied reduction of the surface energy by 1.5\,\%. The available experimental data for Fe as obtained from low-energy electron diffraction on the top-layer relaxation of the bcc (100) surface of Fe is ambiguous with values of $+0.5\,\%$ and $-1.4 \pm 3\%$, both from~\cite{Chiarotti:2012}, and $-5\pm 2\%$ from~\cite{Wang:1987b}. 
For Fe-Cr, the interlayer distances reduce between 2.6$\,\%$ to 2.7$\,\%$ and the surface energies decrease the most for Fe$_{80}$Cr$_{20}$ by 1.4\% and the least for Fe$_{95}$Cr$_{05}$ by 1.2\,\%.
Based on these arguments, we conclude that relaxation is a minor effect in Fe and Fe-Cr and we held fixed all atomic positions to the ideal bcc lattice sites for the results presented in section~\ref{sec:results}.

The total energies as a function of strain, cf.$\!$ \eref{eq:fita} and~\eref{eq:fitb}, were computed in a strain interval of $|\epsilon|=0.02$.

\begin{table}[tbh]
\caption{\label{table:esc}Influence of the choice of the size of the surface subsystem, characterised by $n_{\rm m}$, and the strain interval for elastic energy fits, $|\epsilon|$, for the surface parameters of the (100) surface facet of Fe. The surface parameters of the reference system, given by $n_{\rm m}=13$ and $|\epsilon|=0.02$, are highlighted in boldface. The absolute difference $\Delta$ as well as the relative change in \,\% in parenthesis are specified, and stated at the end of each row  and at the end of each column for a decrease of $|\epsilon|$ from 0.02 to 0.01 and an increase of $n_{\rm m}$ from 13 to 15, respectively. $\gamma$, $\tau$, and the absolute difference are in units of ${\rm J}\cdot{\rm m}^{-2}$. In all cases $n_{\rm v}=7$.}
\begin{indented}
\item[]\begin{tabular}{ccccc}
\br
$n_{\rm m}$  & surface energy $\gamma$ & \multicolumn{2}{c}{surface stress $\tau$}    & $\Delta$ \\
\cline{3-4}
& &  $|\epsilon|=0.02$ & $|\epsilon|=0.01$ & \\
\mr
13 & \bf{2.615} & \bf{0.57} & 0.50 & -0.07 (-12) \\
15 & 2.626 & 0.51 & 0.56 & 0.05 (10) \\
 $\Delta$ & 0.011 (0.4) & -0.06 (-11) & 0.06 (12) & - \\
\br
\end{tabular}
\end{indented}
\end{table}

We conclude this section by establishing the precision of the calculated surface parameters in EMTO with respect to the selected size of the surface subsystem ($n_{\rm m}=13$) and the strain interval ($|\epsilon|=0.02$) using the example of Fe. The previous set of numerical parameters define the reference system, which we consider to yield converged surface related quantities in this work. The surface energy and the surface stress of Fe were computed for a larger surface subsystem ($n_{\rm m}=15$) on the one hand, and the surface stress was fitted to total energies from a narrower strain interval ($|\epsilon|=0.01$) on the other hand. Table~\ref{table:esc} lists both the absolute values of the surface parameters and the change ($\Delta$) with respect to the reference set of surface parameters. Apart from pure Fe, we assessed the precision for several alloy concentrations of Fe-Cr in the same way, and noted that the tabulated $\Delta$'s in table~\ref{table:esc} represent  characteristic values for all concentrations tested. 
$\Delta$ may be used to define the precision of our calculation with respect to the choices of $n_{\rm m}$ and of $|\epsilon|$, which is hence of the order of 0.01-0.02$\,{\rm J}\cdot{\rm m}^{-2}$ for surface energies, and approximately 0.04-0.08$\,{\rm J}\cdot{\rm m}^{-2}$ for surface stresses. The energy scale for surface stress calculations is roughly one order of magnitude smaller than the one for surface energy calculations which explains the difference in the corresponding $\Delta$'s.

Due to the neglect of ASRO in this work, we model a chemically homogeneous bulk alloy, i.e., there is no spatial probability (composition) dependence of the distribution of the alloys components. First we consider an ideally truncated bulk system, i.e., a system with surface, for which no atomic redistribution (segregation) occurs. Hence the surface composition is identical to the one in the bulk for all compositions.
Second, we allow for a change of the surface chemistry.

\section{\label{sec:results}Results and discussion}

\subsection{\label{sec:results:Fe}Surface parameters of Fe}

For the theoretical equilibrium lattice parameter of ferromagnetic (FM) bcc iron, we obtained 2.837\,{\rm \AA} and for the spin moment a value of $2.21\,\mu_{\rm B}$. This is to be compared to experimental values, 2.867\,{\rm \AA}~\cite{Villars:1991} and $2.21\,\mu_{\rm B}$ for the total magnetic moment~\cite{Wijn:1997}.

Table~\ref{table:results:Fe} lists the surface energy and the surface stress of the bcc (100) surface of FM iron as calculated in this work and compared to available data from the literature. The multitude of comparable ab initio data from full-potential (FP) and projector-augmented-wave (PAW) methods allows to draw conclusions on typical scatter of surface energy and surface stress calculations, as well as a critical evaluation of the present results. Concerning the surface energy of Fe, the only outlier seems to be the value obtained with FPLO, since the remaining surface energies scatter in the range of approximately 2.3-2.6$\,{\rm J}\cdot{\rm m}^{-2}$. The particular choice of the gradient corrected density functional may have an effect on the surface energy as all PBE values are larger than the values obtained with the parametrisation of Perdew \etal~\cite{Perdew:1992b,Perdew:1992bE}.
Our EMTO value of $\gamma=2.62\,{\rm J}\cdot{\rm m}^{-2}$ is in close agreement to the PAW and the FP linear augmented plane wave (FP-LAPW) results of Punkkinen \etal.
The too high surface energy from FPLO may be related to a too strongly contracted wave function at the bulk-vacuum interface, cf., e.g., the analysis in~\cite{Fall:1998}. 
Concerning the surface stress of bcc Fe, the FP and PAW values scatters in the range of approximately 1.1-1.4$\,{\rm J}\cdot{\rm m}^{-2}$. The EMTO value, $\tau=0.57\,{\rm J}\cdot{\rm m}^{-2}$, is comparatively small and may thus indicate a systematic underestimation of the surface stress in Fe and the Fe-based system. This underestimation may be ascribed to the muffin-tin approximation to the one-electron potential. 

The experimental value of the surface energy of Fe from~\cite{Tyson:1977}, $\gamma=2.41\,{\rm J}\cdot{\rm m}^{-2}$, is estimated from surface stress measurements of the liquid-vacuum interface at the melting temperature and surface stress measurements of the liquid-solid interface, and extrapolated to $T=0\,{\rm K}$. 

\Table{\label{table:results:Fe}Surface parameters of the bcc (100) surface facet of FM Fe. All methods employed gradient corrected density functionals.}
\br
method &  surface energy $\gamma$ & surface stress $\tau$ & references \\
       & $({\rm J}\cdot{\rm m}^{-2})$ & $({\rm J}\cdot{\rm m}^{-2})$ & \\
\mr
EMTO, PBE &  2.62 & 0.57 & this work \\
 FPLO, PBE &  3.09, 3.07$^a$ &  1.15 & this work \\
 PAW, PBE &  2.55$^b$, 2.50$^a$ & 1.39$^a$ & \cite{Punkkinen:2011b} \\
 FP-LAPW, PBE &  2.6$^b$ & - &  \cite{Punkkinen:2011b} \\
 PAW, GGA\cite{Perdew:1992b,Perdew:1992bE} & 2.48, 2.47$^a$ & - & \cite{Blonski:2007} \\
 PAW, GGA\cite{Perdew:1992b,Perdew:1992bE} & 2.32, 2.29$^a$&- & \cite{Spencer:2002} \\
\cline{1-4}
 experiment  &  2.41$^c$  & - &  \cite{Tyson:1977}\\
\br
\end{tabular}
\item[] $^a$surface layer relaxation included
\item[] $^b$estimated from figure
\item[] $^c$estimated at  $T=0\,{\rm K}$
\end{indented}
\end{table}

Because of the reduced coordination number, the magnetic moment of Fe at the surface is enhanced compared to the bulk. We obtained a surface magnetic moment of $2.97\,\mu_{\rm B}$ in very close agreement with a recently reported value, $2.96\,\mu_{\rm B}$~\cite{Kiejna:2008}. Calculations of Punkkinen \etal~\cite{Punkkinen:2011,Punkkinen:2011b} suggested an almost linear relationship between the magnetic moment enhancement, $\Delta m^2$, and the magnetic surface stress, $\tau_{\rm mag}$, on the basis of their computed surface stresses for the most stable surfaces of magnetically ordered Cr, Mn, Fe, Co, and Ni, i.e.,
\begin{equation}
 \tau_{\rm mag}\propto \Delta m^2 = m^2_{\rm surf}-m^2_{\rm bulk},
\label{eq:taumag}
\end{equation}
where $m_{\rm surf}$ and $m_{\rm bulk}$ are the magnetic moments at the surface and in the bulk, respectively. The magnetic contribution to $\tau$, $\tau_{\rm mag}$, is defined as the difference between the nonmagnetic and the magnetic values of $\tau$, that is evaluated without spin-polarisation ($\tau_{\rm nsp}$) and with spin-polarisation ($\tau_{\rm sp}$) for identical surface geometry, viz. $\tau_{\rm mag}=\tau_{\rm nsp}-\tau_{\rm sp}$. The geometry of the spin-polarised system is the reference state if not stated otherwise.
The above proportionality was verified for elements with FM order (Fe, Co, and Ni) and anti-ferromagnetic (AFM) order (Cr and Mn).
The present EMTO values for FM Fe are $\Delta m^2=3.91\,\mu^2_{\rm B}$ and $\tau_{\rm mag}=3.14\,{\rm J}\cdot{\rm m}^{-2}$, respectively, which fit very well to the correlation established by Punkkinen \etal (see figure~3 from~\cite{Punkkinen:2011}).

\subsection{\label{sec:results:FeCrhom}Chemically homogeneous Fe-Cr alloy}

First we consider the case of chemically homogeneous surface alloys, i.e., it is assumed that the chemical composition at the surface is identical to the bulk composition. 

\subsubsection{\label{sec:results:FeCrhom:latparam}Lattice constants and surface parameters}

The theoretical equilibrium lattice parameters of ferritic Fe-rich Fe-Cr alloys (0-20\,at.\% Cr) were previously calculated with EMTO-CPA and discussed in detail in~\cite{Olsson:2003,Olsson:2006,Zhang:2009,Korzhavyi:2009,Razumovskiy:2011}. Since our computed lattice parameters practically reproduce these earlier results, we refer the reader to those references. It is, however, important to point out the non-linear behaviour of the lattice parameter of the Fe-Cr system.

\begin{figure}
  \resizebox{0.75\columnwidth}{!}{\includegraphics[clip]{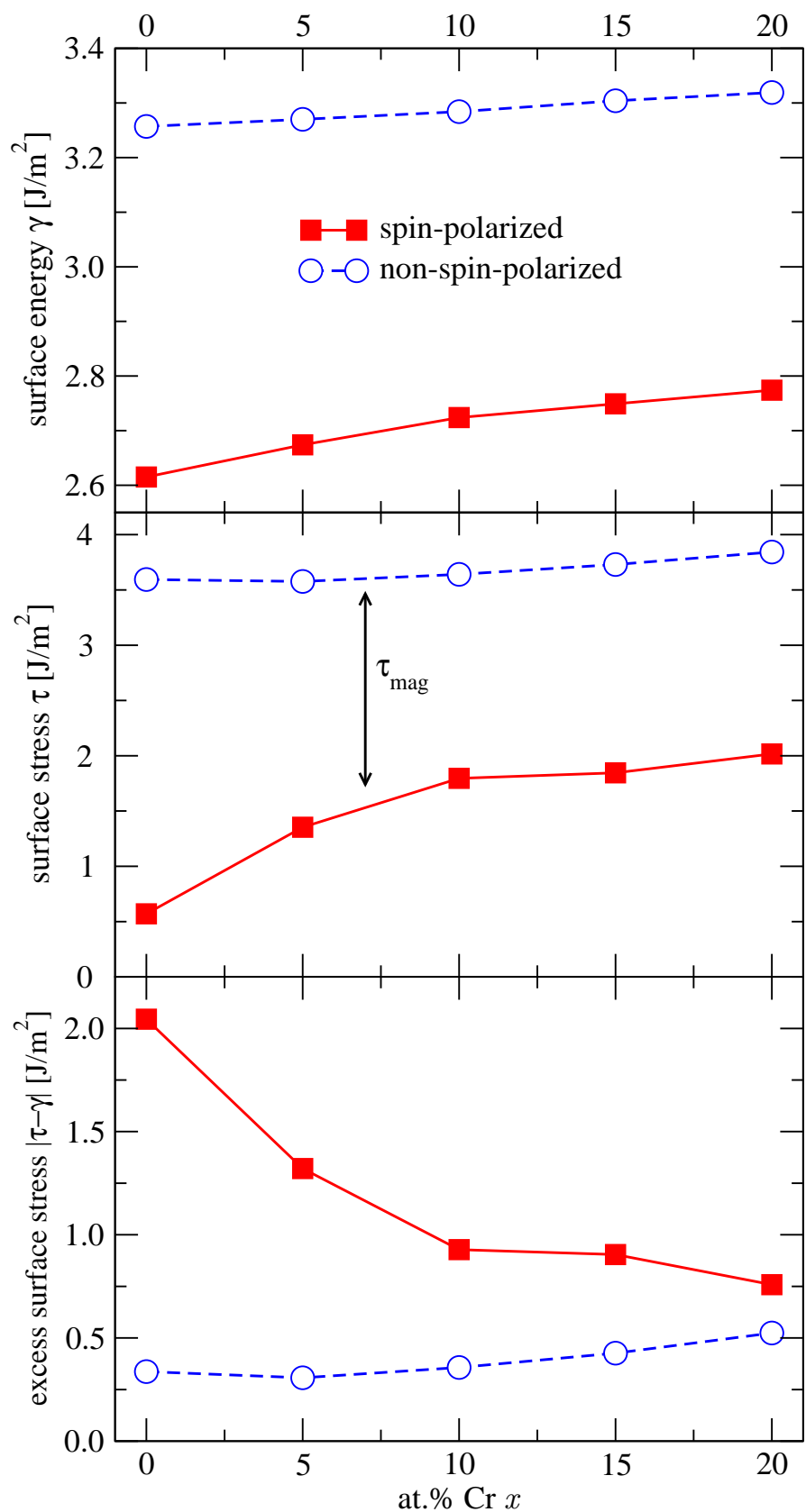}}
\caption{\label{fig:results:FeCrhom:surfparam}Surface parameters (in units of ${\rm J}\cdot{\rm m}^{-2}$) of chemically homogeneous Fe-rich Fe-Cr alloys (0-20\,at.\% Cr): (top panel) surface energy $\gamma$, (middle panel) surface stress $\tau$, (bottom panel) excess surface stress $|\tau-\gamma|$. Lines are a guide to the eye.}
\end{figure}

In the atomic concentration range of 0-20\,at.\% Cr, we identify a clear trend of the concentration dependence of all surface parameters in Fe-Cr alloys, see \fref{fig:results:FeCrhom:surfparam}. 
The surface energy increases monotonically by $0.33\,{\rm J}\cdot{\rm m}^{-2}$ for an increase of the concentration of Cr from $0$ to $20\,{\rm at.}\%$. This trend is not entirely unexpected since the surface energy of the non-relaxed bcc (100) surface facet of Cr was found to be larger by 0.5-0.8$\,{\rm J}\cdot{\rm m}^{-2}$ in theory than the one of Fe.~\cite{Punkkinen:2011b} That is, the surface energy of the disordered alloy with low Cr content (0-20\,at.\% Cr) follows a monotonic trend (rule of mixing) given by the boundary values of pure Fe and pure Cr (the exact PAW values from~\cite{Punkkinen:2011b} for the relaxed surface are $2.50\,{\rm J}\cdot{\rm m}^{-2}$ for Fe and $3.06\,{\rm J}\cdot{\rm m}^{-2}$ for Cr).

In previous theoretical considerations for the most stable surfaces~\cite{Punkkinen:2011,Punkkinen:2011b}, the surface stress of Cr was reported to be $1.9\,{\rm J}\cdot{\rm m}^{-2}$ smaller than the corresponding value of Fe for relaxed surface geometries and likewise was the difference for non-relaxed geometries. As further stated, this difference amounts to $1.7\,{\rm J}\cdot{\rm m}^{-2}$ if specifically the relaxed bcc $(100)$ surface facet is considered (the exact PAW values from~\cite{Punkkinen:2011b} for the relaxed surface are $1.39\,{\rm J}\cdot{\rm m}^{-2}$ for Fe and $-0.32\,{\rm J}\cdot{\rm m}^{-2}$ for Cr). Since relaxation seems to have a similar effect on $\tau$ for both elements for the most stable surface, it is reasonable to assume that the effect of relaxation on $\tau$ for the $(100)$ surfaces of Fe and of Cr are also similar. Hence, the surface stress of the non-relaxed bcc $(100)$ surface of Cr is presumably still considerable smaller than the one of Fe (roughly by 1-2$\,{\rm J}\cdot{\rm m}^{-2}$). 
On this ground, we expect an overall decrease of the surface stress of Fe-Cr with increasing Cr content. Our findings in the dilute Cr concentration range are however contrary to this expectation (\fref{fig:results:FeCrhom:surfparam}). In the range up to $10\,{\rm at.}\%$ Cr in the iron matrix, the surface stress strongly increases to a maximum value of $1.78\,{\rm J}\cdot{\rm m}^{-2}$ being approximately $1.25\,{\rm J}\cdot{\rm m}^{-2}$ larger than the corresponding value of Fe. $\tau$ levels off for concentrations higher than $10\,{\rm at.}\%$ Cr. 

The third surface-characteristic quantity depicted in \fref{fig:results:FeCrhom:surfparam} is the excess surface stress, $|\tau-\gamma|$. It evidently decreases strongly in the concentration range up to $10\,{\rm at.}\%$ Cr relative to the value of pure Fe, which is mainly due to the accompanied increase of $\tau$. For Cr concentrations above $10\,{\rm at.}\%$,  $|\tau-\gamma|$ remains almost unchanged. The surface reconstruction is predicted to occur when the excess surface stress becomes larger than the characteristic surface strain energy associated with the reconstruction~\cite{Cammarata:1992,Zolyomi:2009}. The latter may be expressed in terms of the shear modulus and the Burgers vector. Now, taking into account that the elastic moduli of Fe-Cr alloys show a rather weak composition dependence for the present concentration interval~\cite{Zhang:2009}, one may conclude, that the $(100)$ surface of Fe-rich Fe-Cr alloys is considerable more stable against reconstruction than the surface of pure Fe.

\subsubsection{Magnetism and magnetic surface stress}

The magnetic structure of ferritic Fe-Cr is governed by interactions between Fe atoms, that prefer to align their magnetic moments in parallel, and Cr atoms, that favour an anti-parallel alignment. Its energetics is rather well described within collinear magnetism of fixed Ising spins~\cite{Ackland:2006,Olsson:2006,Korzhavyi:2009}. 
In the iron-rich ferritic Fe-Cr alloys, the moment at Cr sites are coupled anti-parallel to the ones of Fe necessarily implying that they are aligned in unfavourable parallel orientation with respect to other Cr atoms. In the dilute limit, however, the average Cr-Cr distance is large and their mutual interaction energy small~\cite{Ropo:2011,Klaver:2006}. 
The EMTO-CPA spin moment at Cr is $-1.62\,\mu_{\rm B}$ on the impurity level. This calculation was done with $0.05\,{\rm at.}\%$ Cr. A negative sign of the magnetic moment indicates an antiparallel alignment with respect to the moment of Fe which was defined to possess a positive sign. 
Klaver \etal~obtained a slightly larger spin moment of Cr in the dilute limit, $-1.8\,\mu_{\rm B}$, calculated with the PAW method and PBE for a super cell with an effective Cr concentration of $0.19\,{\rm at.}\%$~\cite{Klaver:2006}.
Thus, a Cr atom in a surrounding Fe matrix at very low Cr concentrations is much stronger polarised than in AFM ordered pure Cr (the experimental spin moment of the long wave spin-density ground state of Cr is approximately $0.59\,\mu_{\rm B}$).

\begin{figure}
  \resizebox{0.9\columnwidth}{!}{\includegraphics[clip]{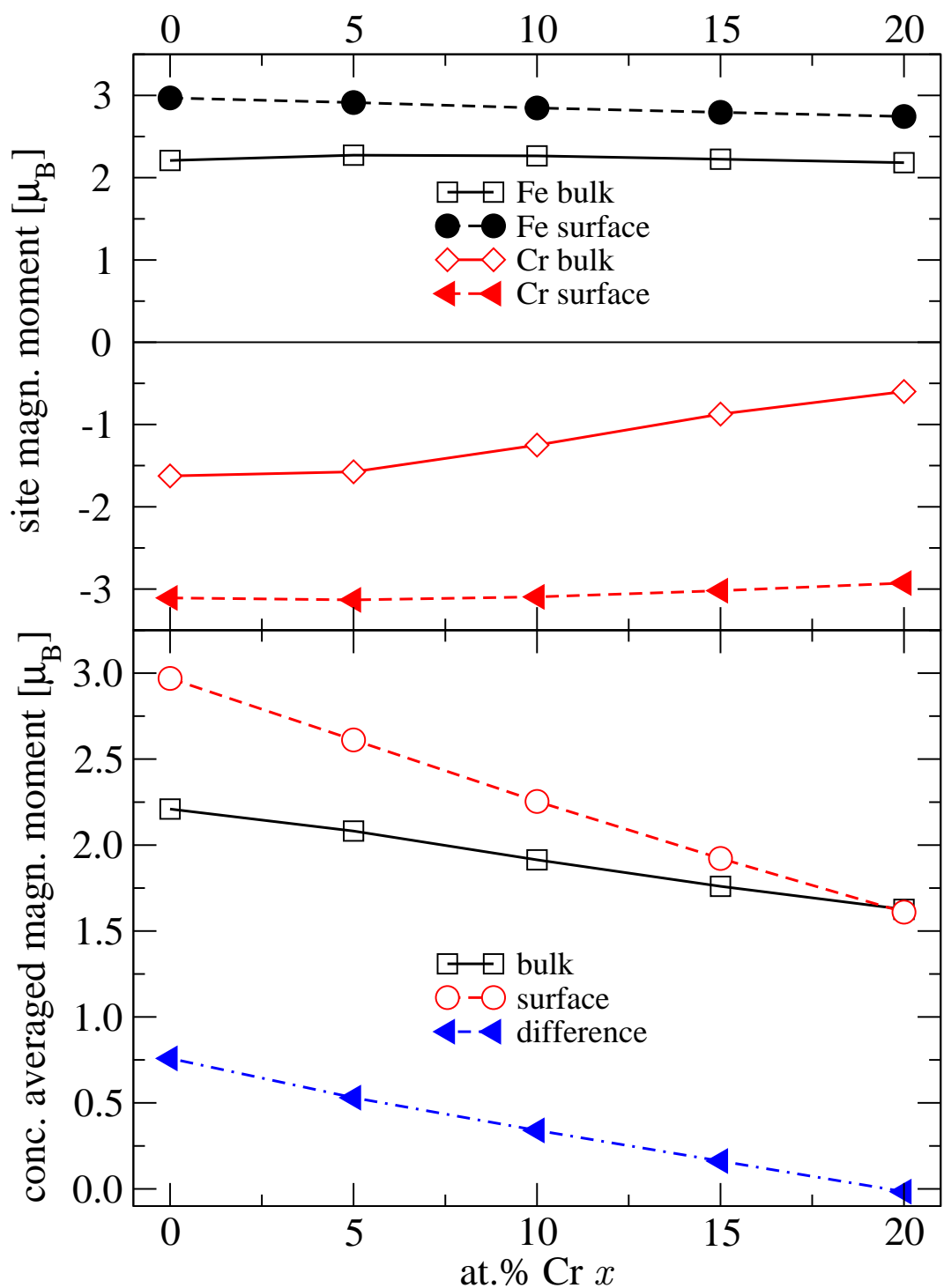}}
\caption{\label{fig:results:FeCrhom:magnetism}Magnetism in Fe-Cr alloys for the chemically homogeneous surface and bulk reference systems. Top panel: site resolved spin magnetic moments in the bulk and at the surface; bottom panel: concentration averaged spin magnetic moments in the bulk and at the surface, and the difference between surface and bulk magnetic moments. 
EMTO-CPA calculations on the impurity level were done for $0.05\,{\rm at.}\%$ Cr. Lines are a guide to the eye.}
\end{figure}

Within the random solid solution description provided by EMTO-CPA, increasing the Cr concentration beyond dilute levels results in a gradual loss of the modulus of the magnetic moment at Cr sites while the Fe spin moment hardly changes, see \fref{fig:results:FeCrhom:magnetism}. These findings are in line with previously published theoretical assessments within the CPA~\cite{Olsson:2006,Ruban:2008}. This concentration dependent effect on the Cr moments seems to be well understood on the basis of an increased number of unfavourable Cr-Cr interactions (frustration) with increasing number of Cr atoms in the Fe matrix.
The total net magnetic moment of the alloys decreases in the same concentration interval (\fref{fig:results:FeCrhom:magnetism}). 
As reported in~\cite{Olsson:2006}, the total net magnetic moment of Fe-Cr obtained in the CPA is in close agreement with the measured net magnetic moments in the FM phase of Fe-Cr. 
Klaver \etal~and Korzhavyi \etal~showed by means of the super cell technique, that clustering of Cr atoms in Fe-Cr in the concentration range $\le 20\,{\rm at.}\%$ Cr leads to a reduction of the absolute value of the Cr magnetic moments in comparison to dispersed Cr atoms due to frustration~\cite{Klaver:2006,Korzhavyi:2009}. As mentioned above, the onset of clustering of Cr atoms can be connected to the experimentally determined inversion of the ASRO parameter at approximately $6\,{\rm at.}\%$ Cr~\cite{Dubiel:2011,Idczak:2012} or at approximately $11\,{\rm at.}\%$ Cr~\cite{Mirebeau:1984,Mirebeau:2010}.

Both the magnetic moments of Fe atoms and of Cr atoms located at the surface are enhanced with respect to their bulk values. The EMTO-CPA magnetic moment of Cr located at the surface in the dilute limit amounts to $-3.11\,\mu_{\rm B}$ being thus even larger in absolute value than the corresponding value of Fe ($2.96\,\mu_{\rm B}$).
We realise from \fref{fig:results:FeCrhom:magnetism} that Fe moments in the surface layer and in the bulk change little as a function of concentration in contrast to the magnetism at the Cr sites. The modulus of the spin moment of a Cr atom localised at the surface undergoes a slight decrease in the range of increasing Cr content from 0 to 20\,at.\%, which is in fact similar to the reduction of the Fe surface moment. The different behaviours of the Cr-Cr interaction is due to the different average distance between two Cr atoms at the surface and in the bulk (for the same concentration it is larger at the surface) and the interaction energy which scales with the number of atoms in nearest neighbour shells (which is larger in the bulk)~\cite{Ropo:2011}.
As discussed above, the absolute value of the bulk Cr moment drops considerably resulting in a drastically higher moment enhancement of Cr at the surface.
This strong moment enhancement is in fact a propensity of atoms in the surface layer only; Fe and Cr magnetic moments in subsurface layers of the bcc (100) surface possess almost bulk values~\cite{Ropo:2011,Kiejna:2008}.

As a consequence of the distinct moment behaviours of the individual alloys components, the total net surface moment of the alloy diminishes more pronounced than the total net bulk moment (\fref{fig:results:FeCrhom:magnetism}). Their difference drops to zero at approximately $20\,{\rm at.}\%$ Cr, i.e., the net surface moments equals the net bulk moment. We suggest that the changes in the magnetic structure determine the trends of $\gamma$ and $\tau$ as we argue below.

To understand the contribution of magnetism to the noticed behaviour of surface parameters we return to \fref{fig:results:FeCrhom:surfparam} where we included data of non-spin-polarised calculations as well. These were done for exactly the same geometry as the spin-polarised calculations. The magnetic contribution to the surface energy, $\gamma_{\rm mag}$, is likewise defined to $\tau_{\rm mag}$ as the difference between the nonmagnetic and the magnetic values of $\gamma$. In agreement with previous investigations for $3d$ transition metals~\cite{Punkkinen:2011,Alden:1992}, we find that magnetism generally reduces surface energies and surface stresses in the case of Fe-Cr. The magnitude of the magnetic contribution to the stress is clearly larger than its effect on the surface energy. Furthermore, $\tau_{\rm mag}$ reduces strongly by approximately 1.3$\,{\rm J}\cdot{\rm m}^{-2}$ in the concentration range from 0 to 10\,at.\% Cr while it is approximately constant for higher concentrations. $\gamma_{\rm mag}$ exhibits the same behaviour, however less pronounced. These behaviours confirm that changes in the magnetic structure drive the observed trends of $\gamma$ and $\tau$.

\begin{figure}
  \resizebox{0.95\columnwidth}{!}{\includegraphics[clip]{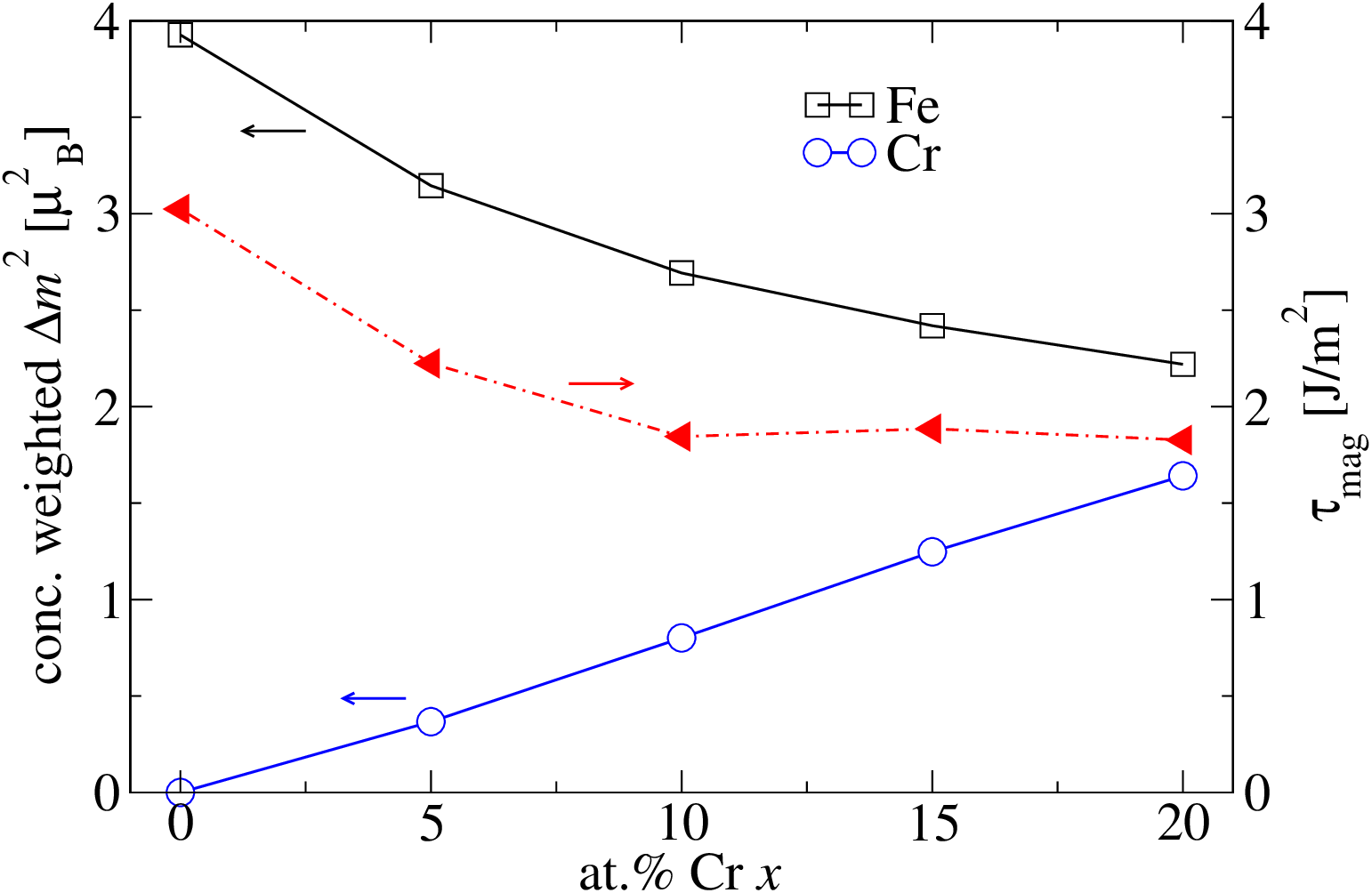}}
\caption{\label{fig:results:FeCrhom:tau_mag_mom_enh}Contributions of Fe and Cr to the surface moment enhancement, $\Delta m^2$, weighted by their atomic concentration (left-hand ordinate) and magnetic surface stress $\tau_{\rm mag}$ (right-hand ordinate) for the chemically homogeneous reference systems. Note that $\tau_{\rm mag}$ equals the indicated $\tau_{\rm mag}$ from \fref{fig:results:FeCrhom:surfparam}. Lines are a guide to the eye.}
\end{figure}

Both Fe and Cr are polarised in Fe-Cr, hence both species contribute to $\tau_{\rm mag}$. To better understand their individual contributions, we explore the concentration weighted surface moment enhancement, that is, we evaluate \eref{eq:taumag} separately for Fe and Cr and weight the results by $(1-x)$ and $x$, respectively. The resulting data in \fref{fig:results:FeCrhom:tau_mag_mom_enh} signals a correlation between $(1-x)\Delta m^2 ({\rm Fe})$ and $\tau_{\rm mag}$ in the concentration range $x\le 20\,{\rm at.}\%$ Cr. The effect of Cr on $\tau_{\rm mag}$ is strongly diminished since the monotonically increasing weighted surface moment enhancement of Cr, $x\Delta m^2({\rm Cr})$, is not reflected in the trend of $\tau_{\rm mag}$. Chromium may, however, be associated with the levelling-off of the magnetic surface stress for Cr concentrations in the range $10\,\% \le x\le 20\,\%$. Thus, in Fe-rich Fe-Cr alloys the trend of $\tau_{\rm mag}$ as a function of concentration seems to be dominated by the magnetism of Fe.

\subsubsection{Magnetic pressure}

Starting from the definition of the magnetic surface stress ($\tau_{\rm mag}=\tau_{\rm nsp}-\tau_{\rm sp}$) and using \eref{eq:surfstress:fit}, we regroup all appearing terms according to
\begin{eqnarray}
 \tau_{\rm mag} &=& \frac{c^{\rm nsp}_{\rm surf}-c^{\rm nsp}_{\rm bulk}}{4A} - \frac{c^{\rm sp}_{\rm surf}-c^{\rm sp}_{\rm bulk}}{4A} \nonumber \\
&=& {} \frac{c^{\rm nsp}_{\rm surf}-c^{\rm sp}_{\rm surf}}{4A} - \frac{c^{\rm nsp}_{\rm bulk}-c^{\rm sp}_{\rm bulk}}{4A} \nonumber \\
&\equiv& {} \tau^{\rm surf}_{\rm mag} - \tau^{\rm bulk}_{\rm mag}.
\end{eqnarray}
In the previous line, we defined the magnetic stress of the surface reference system, $\tau^{\rm surf}_{\rm mag}$, and the magnetic stress of the bulk reference system,  $\tau^{\rm bulk}_{\rm mag}$. Notice that the definition of $\tau^{\rm bulk}_{\rm mag}$ does not include any parameter related to surfaces anymore. It quantifies how much magnetism contributes to the bulk stress and it is thus closely related to the magnetic pressure known since the advent of band structure theory~\cite{Andersen:1985,Andersen:1977}. Magnetic pressure is for example associated with increased atomic volumes in FM transition metals compared to what their volumes would be in the absence of spin-polarisation. 

\begin{figure}
  \resizebox{0.95\columnwidth}{!}{\includegraphics[clip]{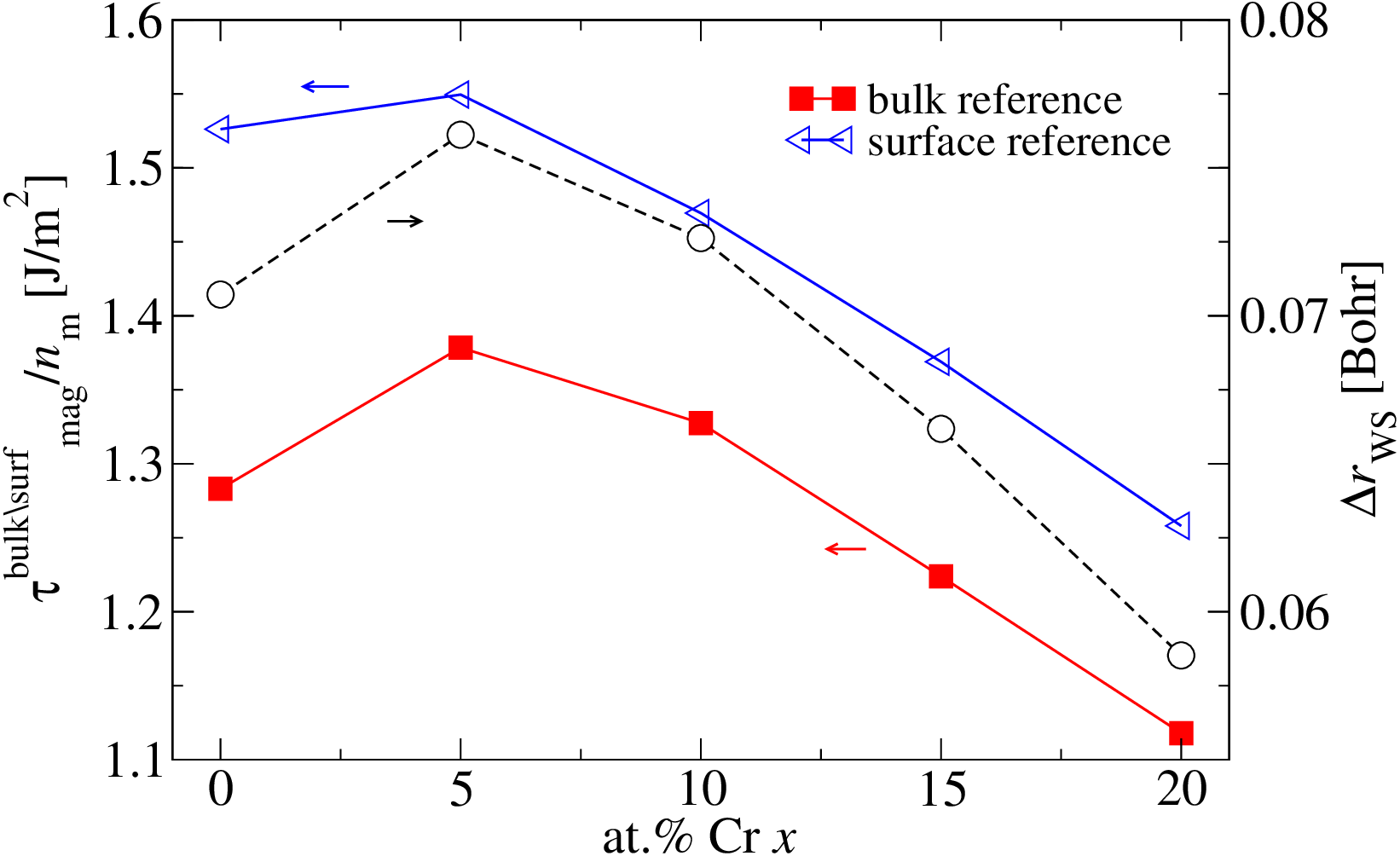}}
\caption{\label{fig:results:FeCrhom:tau_bulk_diff_latparam}Correlation between the magnetic stress of the chemically homogeneous bulk and surface reference systems per layer and per particle (left-hand ordinate) and the difference in the equilibrium Wigner-Seitz radius between the FM Fe-Cr system and the non-polarised Fe-Cr system (right-hand ordinate). Lines are a guide to the eye.}
\end{figure}

The magnetic pressure of the bulk reference system per layer and per particle, $\tau^{\rm bulk}_{\rm mag}/n_{\rm m}$, as a function of concentration for the Fe-Cr system as plotted in \fref{fig:results:FeCrhom:tau_bulk_diff_latparam} is non-linear and tensile, indicating that without magnetism the lattice constant, or equivalently, the Wigner-Seitz radius would be smaller. The order of $\tau^{\rm bulk}_{\rm mag}$ may be connected to the difference in the Wigner-Seitz radius, $r_{\rm WS}$, between the FM Fe-Cr system and the nonmagnetic Fe-Cr system.  
Our data shows that the equilibrium Wigner-Seitz radius of the nonmagnetic Fe-Cr system follows a linear concentration dependence, i.e, Vegard's rule ($ r_{\rm WS}({\rm Fe}_{1-x}{\rm Cr}_{x})\sim(1-x)r_{\rm WS}({\rm Fe})+xr_{\rm WS}({\rm Cr}$)), with $r_{\rm WS}({\rm Cr})> r_{\rm WS}({\rm Fe})$. As mentioned in the beginning of Sec.~\ref{sec:results:FeCrhom:latparam}, the equilibrium Wigner-Seitz radius of the FM Fe-Cr system changes non-linearly as a function of the Cr concentration. The difference in the equilibrium Wigner-Seitz radius between the FM Fe-Cr alloy and the NM Fe-Cr alloy, $\Delta r_{\rm WS}$, is plotted in \fref{fig:results:FeCrhom:tau_bulk_diff_latparam}. We find that $\tau^{\rm bulk}_{\rm mag}$ and $\Delta r_{\rm WS}$ are strongly correlated: $\tau^{\rm bulk}_{\rm mag}$ as a quantitative measure for the magnetic pressure in the bulk Fe-Cr system correlates with the deviation of the equilibrium Wigner-Seitz radius from Vegard's rule.

The magnetic pressure of the surface reference system ($\tau^{\rm surf}_{\rm mag}$) tells us about the effect of magnetism at the surface on $\tau$ compared to a magnetism-free surface. The surface reference systems, however, include bulk-like contributions, as effects due to the presence of a surface decay towards the interior of the surface reference systems which thus become gradually more bulk-like as the distance to the surface is increased. Therefore $\tau^{\rm surf}_{\rm mag}$ and $\Delta r_{\rm WS}$ are also correlated in the same way as $\tau^{\rm bulk}_{\rm mag}$ and $\Delta r_{\rm WS}$ are, see the plot of $\tau^{\rm surf}_{\rm mag}/n_{\rm m}$ in \fref{fig:results:FeCrhom:tau_bulk_diff_latparam}. The differences between $\tau^{\rm surf}_{\rm mag}$ and $\tau^{\rm bulk}_{\rm mag}$, i.e., both the absolute value and the trend as a function of $x$, are ascribed to both the spin-polarised and the non-spin-polarised surfaces. It is interesting to note that the maximum of $\tau^{\rm surf}_{\rm mag}$ seems to be well below 5\,at.\% Cr while it is above 5\,at.\% Cr for $\tau^{\rm bulk}_{\rm mag}$. 

Magnetism is the driving force for the enlarged lattice parameters of the FM Fe-Cr system compared to the non-polarised model system. Magnetism leads on the other hand to increased magnetic stresses of the bulk and the surface systems. Magnetic stresses are compressive, i.e., have a tendency to expand the lattice. According to \fref{fig:results:FeCrhom:tau_bulk_diff_latparam}, $\tau^{\rm surf}_{\rm mag}$ is larger than $\tau^{\rm bulk}_{\rm mag}$, i.e., the magnetic contributions to $\tau$ favour a larger surface lattice parameter than in the bulk.

\subsection{\label{sec:results:FeCrinhom}Chemically inhomogeneous Fe-Cr alloy}

There is experimental and theoretical evidence for segregation on Fe-Cr surfaces. According to first-principles calculations of Ropo \etal~for the equilibrium segregation profile of the (100) surface at various temperatures, the surface chemistry seems to be determined by the bulk configuration which leads to the peculiar threshold behaviour at approximately 10\,at.\% Cr bulk content~\cite{Ropo:2007,Ropo:2011}. Below this value, the thermodynamically most stable (100) surface was found to be essentially Fe-terminated in agreement with first-principles surface segregation calculations of Cr in Fe-rich Fe-Cr~\cite{Ponomareva:2007}, while this surface facet is enriched in Cr to even higher than bulk concentrations above this threshold up to approximately 17\,at.\% Cr in the bulk. Owning the complexity of these calculations only the surface layer concentration was variable.

Since one nevertheless may expect that the (magnetic) contributions to both excess quantities originate from the topmost layers on every surface, we have good reasons to believe that changing only the Cr concentration of the surface layer captures the dominant effect on the trends of both surface energy and surface stress with surface alloying. 
It is then of course of interest to track how these surface parameters behave and accordingly how the stability of the surface is affected. 

The surface energies for chemically inhomogeneous concentration profiles $\{x_\alpha\}$ of binary A$_{1-x}$B$_{x}$ alloys are obtained according to~\cite{Ruban:1999,Pourovskii:2003},
\begin{eqnarray}
\gamma(\{x_\alpha\}) &=&  \frac{E^n_{\rm surf}(\{x_\alpha\})-\frac{n_{\rm m}}{n}E^n_{\rm bulk}(x)}{2A} \nonumber  \\
 {} & & - \frac{ \Delta\mu_{\rm bulk}(x) \sum_{\alpha=1}^{n_m}(x_{\alpha}-x) }{2A},
 \label{eq:surfaceenergy:slabs:inhom}
\end{eqnarray}
where $\{x_\alpha\}=x_1,x_2,\ldots,x_{n_{\rm m}}$ denotes the concentration of the B element within the layer $\alpha$ perpendicular to the surface, $x_1$ and $x_{n_{\rm m}}$ being the concentrations of the two surface layers and $x_i=x_{n_{\rm m}-i+1}$ due to the symmetry of the slab. The bulk effective chemical potential, $\Delta\mu_{\rm bulk}(x)$, equals the difference of the chemical potentials of the two alloy components in the bulk and is derived from the bulk energy (per atom), viz.
\begin{equation}
 \Delta\mu_{\rm bulk}(x) = \frac{{\rm d} E_{\rm bulk}(x) }{{\rm d} x}.
\end{equation}
The surface stress of the bcc (100) surface facet for a chemically inhomogeneous concentration profile is then obtained by, using \eref{eq:surfstress:bcc100} for the $n$-layer slab,
\begin{eqnarray}
\tau(\{x_\alpha\}) &=& \left.\frac{1}{4A}\Bigg(\frac{\partial (E^{n}_{\rm surf} - \frac{n_{\rm m}}{n}E^{n}_{\rm bulk})}{\partial \epsilon}\right|_{\epsilon=0} \label{eq:surfstress:bcc100:inhom} \\
{} & & {} -\left. \frac{\partial  \left( \Delta\mu_{\rm bulk}(x) \sum_{\alpha=1}^{n_m}(x_{\alpha}-x) \right)  }{\partial \epsilon} \right|_{\epsilon=0}\Bigg). \nonumber
\end{eqnarray}

\begin{figure}[htbp]
 \begin{center}
   \resizebox{0.95\columnwidth}{!}{\includegraphics[clip]{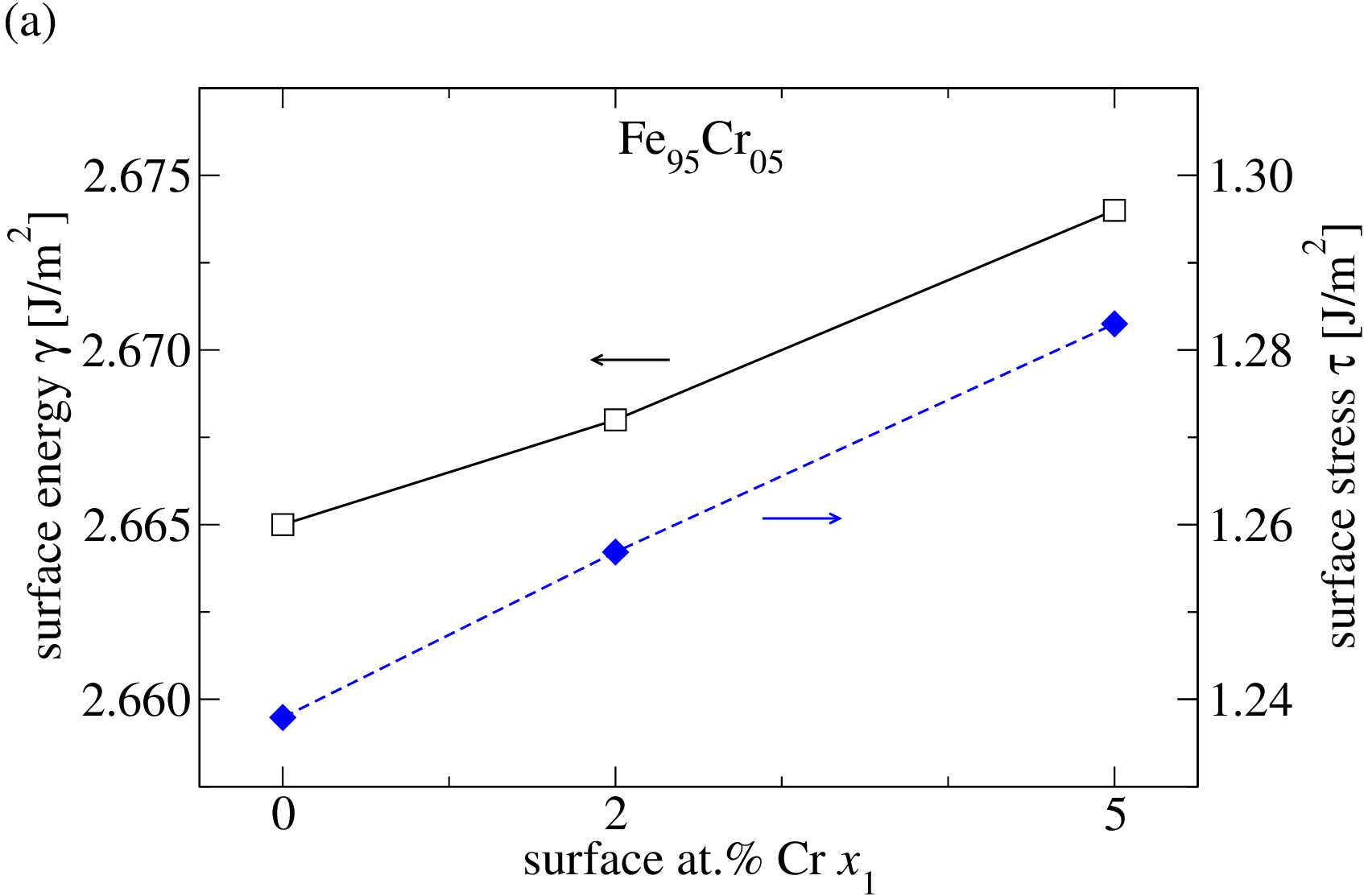}} \\
   \resizebox{0.95\columnwidth}{!}{\includegraphics[clip]{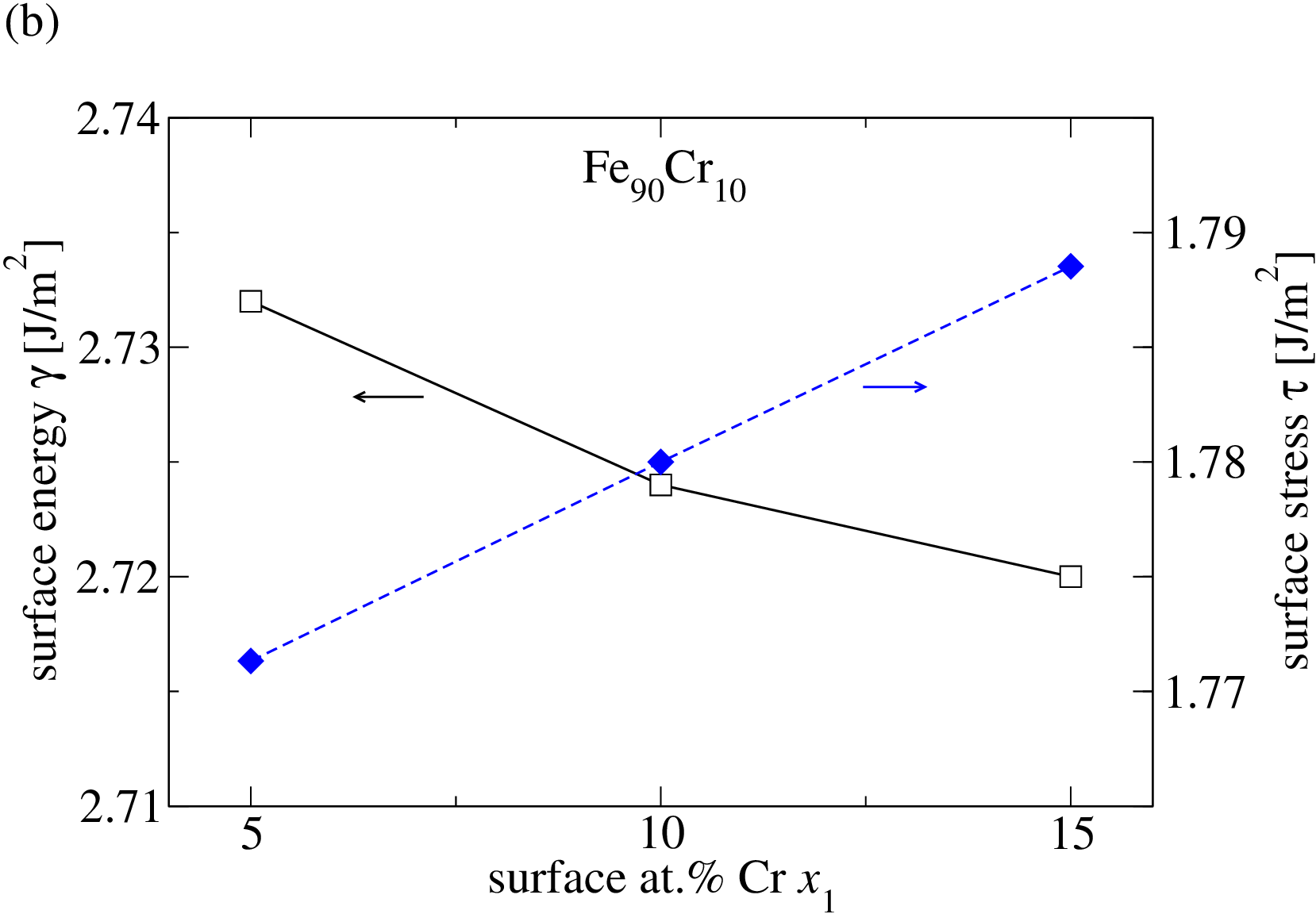}} 
 \end{center}
\caption{\label{fig:chap:surf:alloy:surfparam:inhomo:surfparam}Surface parameters of Fe-rich Fe-Cr alloys with chemically inhomogeneous concentration profiles for two distinct bulk alloy systems, (a) Fe$_{95}$Cr$_{05}$ and (b) Fe$_{90}$Cr$_{10}$, as a function of the concentration of Cr in the surface layer, $x_1$. Lines are a guide to the eye.}
\end{figure}

In this work, we are mostly interested in identifying trends that arise if the homogeneous concentration profile is altered towards a chemically inhomogeneous profile. Here we were guided by the aforementioned computed equilibrium profile at $T=0\,{\rm K}$ for the bcc $(100)$ surface of Fe-Cr as reported by Ropo \etal~\cite{Ropo:2007,Ropo:2011}, i.e., the concentration profiles are variable in the surface layer concentration, $x_1$ ($x_1=x_{n_{\rm m}}$), and all other concentrations are fixed to the bulk value, $x_2=x_3=\ldots=x_{n_{\rm m}-1}=x$. 
Two distinct bulk alloy systems, Fe$_{95}$Cr$_{05}$ and Fe$_{90}$Cr$_{10}$, were selected. The former has a predicted equilibrium surface concentration of 0\,at.\% Cr, i.e., a surface Cr concentration lower than the bulk one, and the latter possesses a predicted equilibrium surface concentration of 15\,at.\% Cr, i.e., a surface Cr concentration larger than the bulk one. For the Fe$_{95}$Cr$_{05}$ bulk alloy system, we thus monitored $\gamma$ and $\tau$ with gradual reduction of the Cr surface concentration from 5 to 0\,at.\%. 
For the Fe$_{90}$Cr$_{10}$ bulk alloy we varied the Cr amount at the surface between 5 and 15\,at.\%.

The effective chemical potential depends on the strain \eref{eq:surfstress:bcc100:inhom}. Applying a strain of the size $\epsilon =\pm 0.02$ led to a relative change of $|\Delta\mu_{\rm bulk}(x)|$ of approximately $10^{-5}$. Since $\sum_{\alpha=1}^{n_m}(x_{\alpha}-x)=2(x_{1}-x)$ is at most 0.1 in the present work, the second term in \eref{eq:surfstress:bcc100:inhom} containing the effective chemical potential is approximately by a factor of $10^{-3}$ smaller than the first one for the current system. 
Thus, we assume that the strain dependence of the effective chemical potential can be neglected, and $\tau$ is again obtained from \eref{eq:surfstress:fit}. Thus, the coefficient $c_{\rm surf}$ is the one for the chemically homogeneous system with fixed bulk concentration $x$. This seems reasonable since the surface chemistry should not affect the bulk contributions to the surface stress. We recall that $c_{\rm surf}$ should be identically zero at bulk equilibrium.

The surface parameters for the two different bulk alloy systems are compiled in \fref{fig:chap:surf:alloy:surfparam:inhomo:surfparam}. 
For the Fe$_{95}$Cr$_{05}$ bulk alloy, we identify the following trend for the spin-polarised calculations: less surface Cr reduces both the surface energy and the surface stress, i.e., the Cr free surface possesses the lowest surface energy and the lowest surface stress. The excess surface stress (not shown) increases slightly by 0.02$\,{\rm J}\cdot{\rm m}^{-2}$ when the surface Cr concentration is reduced from 5\,at.\% to 0\,at.\%.
The bulk alloys containing 10\,at.\% Cr show different behaviours: the richer the surface in Cr is the smaller is the surface energy and the larger is the surface stress. As a result of these trends, the excess surface stress of the surface with $x_1=$15\,at.\% Cr is by 0.03$\,{\rm J}\cdot{\rm m}^{-2}$ more stable than the low-Cr surface.

Our data show that an Fe-rich surface has the lowest surface energy for a bulk concentration below the anticipated threshold concentration, while above this threshold a Cr-enriched surface possesses the lowest surface energy. This finding is in qualitative agreement with calculations for the surface energy of the bcc (100) facet of Fe-Cr from~\cite{Ropo:2011} done for fixed $x_1=\{0,10\}$ at.\% Cr and variable bulk concentration. Recalling that the global trend of $\gamma$ as a function of the bulk Cr concentration shows a homogeneously increasing tendency in $x$ (\fref{fig:results:FeCrhom:surfparam}), we realise that changing the surface composition may alter this picture, as in the case of the Fe$_{90}$Cr$_{10}$ alloy.

We find for the investigated inhomogeneous surfaces, that a larger amount of Cr in the bulk and at the surface increases the surface stress $\tau$ and decreases the magnetic surface stress $\tau_{\rm mag}$ (data for the inhomogeneous surface is not shown). That is, Cr addition drives the tendency towards smaller in-plane lattice constants at the surface compared to the bulk, and Cr addition has a tendency to reduce the compressive magnetic contribution to the total surface stress.

\section{Conclusion}

Using the EMTO method and the CPA we computed the surface energy, the surface stress, and the excess surface stress of the thermodynamically most stable surface facet (100) of the homogeneous disordered bcc Fe-Cr system in the concentration interval up to 20 at.\% Cr. We found that the surface energy increases monotonically with Cr addition thereby following the rule of mixing. 
An increase of the bulk amount of Cr also increases the surface stress, which is unexpected, since the surface stress of Cr is considerably smaller than the one of Fe. As a result of this surprising trend, the excess surface stress reduces with increasing Cr concentration meaning that the (100) surface of Fe-Cr becomes more stable against reconstruction than the same surface of Fe. 

The reduction of the compressive magnetic contribution to the total surface stress (magnetic surface stress) was identified to dominate this increase of the surface stress. We showed further that mainly the magnetic moment enhancement of Fe is correlated with the behaviour of the magnetic surface stress. Thus, we conclude that mainly the magnetism of Fe in Fe-Cr up to 20 at.\% Cr is responsible for the unexpected trend of the surface stress.
Since the surface stress of pure Cr low-index surface facets is much smaller than the one of pure Fe (which was previously shown to be due to magnetism), we expect that Cr replaces Fe to dominate the magnetic surface stress for larger concentrations than 20 at.\% Cr.

We also investigated the impact of surface segregation on the surface parameters for the Fe$_{95}$Cr$_{05}$ and Fe$_{90}$Cr$_{10}$ alloys. The former was previously shown to be Fe-terminated while the latter was shown to be enriched in Cr in vacuum~\cite{Ropo:2007,Ropo:2011}. Varying only the concentration of the surface layer, we established the following trends: a larger amount of surface Cr increases the surface stress for both systems, while Cr addition raises (lowers) the surface energy for the bulk Fe$_{95}$Cr$_{05}$ (Fe$_{90}$Cr$_{10}$) alloy. 

For all investigated chemically homogeneous and inhomogeneous disordered surface profiles, a larger amount of Cr in the alloy favours a smaller in-plane lattice constants at the surface than in the bulk, and the addition of Cr shows a tendency to reduce the compressive magnetic contribution to the total surface stress. 

\ack
The Swedish Research Council, the Swedish Steel Producers' Association, the European Research Council, and the Hungarian Scientific Research Fund (research project OTKA 84078) are acknowledged for financial support. S.~S. gratefully acknowledges the Carl Tryggers Stiftelse f\"or Vetenskaplig Forskning and the Olle Erikssons Stiftelse f\"or Materialteknik.

\end{document}